\documentclass{aastex}

\shorttitle{Atomic Carbon in Galactic Center}
\shortauthors{Ojha et al.}

\begin{document}

\title{AST/RO Observations of Atomic Carbon near the Galactic Center }

\author{Roopesh Ojha\altaffilmark{1}, Antony A. Stark\altaffilmark{1},
Henry H. Hsieh\altaffilmark{1}, Adair P. Lane\altaffilmark{1}}
\affil{Harvard-Smithsonian Center for Astrophysics}
\affil{60 Garden St., MS 78; Cambridge MA 02138} 
\author{Richard A. Chamberlin\altaffilmark{2},}
\affil{Caltech Submillimeter Obs., 111 Nowelo Street }
\affil{Hilo, HI 96720}
\author{Thomas M. Bania\altaffilmark{3}, Alberto D. Bolatto\altaffilmark{3},
 James M. Jackson\altaffilmark{3},}
\affil{Astronomy Department, Boston University, 725 Commonwealth Avenue}
\affil{Boston, MA 02215 }
\and
\author{Gregory A. Wright\altaffilmark{4} }
\affil{Bell Laboratories }
\affil{Holmdel, NJ 07733}
\altaffiltext{1}{rojha,aas,hhsieh,alane@cfa.harvard.edu}
\altaffiltext{2}{cham@ulu.submm.caltech.edu}
\altaffiltext{3}{bania,bolatto,jackson@bu.edu}
\altaffiltext{4}{gwright@bell-labs.com}

\begin{abstract}
We present a coarsely-sampled map of the region $|l|\leq2^\circ$,
$|b|\leq$ $0.1^\circ$ in the 492 GHz
$(^{3}P_{1}\rightarrow^{3}P_{0})$ fine structure 
transition of neutral carbon, observed with the Antarctic
Submillimeter Telescope and Remote Observatory (AST/RO). The
distribution of [\ion{C}{1}] emission is similar on the large scale to that
of CO $J=1\rightarrow0$. On average, the ratio of the integrated
intensities, $\rm{I}_{[\rm{C I}]}/\rm{I}_{^{12}CO}$, is higher in the
Galactic disk than in the Galactic Center region. This result is
accounted for by the absorption of $^{12}\rm{CO}$ within the clouds 
located in the outer Galactic disk.  
$\rm{I}_{[\rm{C I}]}/\rm{I}_{^{12}CO}$ is surprisingly 
uniform over the variety of environments near the Galactic Center. On
average, [\ion{C}{1}] is optically thin (or as optically thin as
$^{13}\rm{CO}$ $(J=1\rightarrow0)$) even in the dense molecular clouds of
the Galactic Center region.  
\end{abstract}


\section{Introduction}
\label{s:intro}

The Galactic Center is a complex region displaying an array of
interesting astrophysical phenomena, many of which indicate an
environment very different from the Galactic disk (see Morris \&
Serabyn 1996 for a recent review). One striking feature is that the
average density of molecular material is much higher in the Galactic
Center than anywhere else in the Galaxy, with the Galactic Center ISM
dominated by molecular cloud complexes. Though the inner 500 pc of the
Galaxy is less than $0.2\%$, by volume, of the galactic disk it
contains nearly $10\%$ of the Galaxy's molecular mass.  Galactic
Center molecular clouds are distinctly different from molecular clouds
in the disk. They have large internal velocity dispersions with line
widths 4 to 5 times wider than is typical of the galactic disk.
They are warm and have a high density, which lets them survive the
tidal disruptions of the central mass concentration (Stark et al. 1991).
 
The fine structure lines of neutral atomic carbon $\rm{C}^{0}$ are an
important source of information about the physics and chemistry of
molecular clouds, and play an important role in their energy
balance. These lines are predicted to be the primary coolants
throughout the partially opaque ($A_{v}\sim$1-3) regions of
molecular clouds (Wolfire et al. 1995). The fine structure [\ion{C}{1}]
$(^{3}P_{1}\rightarrow^{3}P_{0})$ transition of neutral carbon is
excited at densities $n\gtrsim500$ cm$^{-3}$ and temperatures
$T\gtrsim24$~K.  It traces phases of the molecular ISM where CO will
either not be excited (the critical density for the CO
$J=1\rightarrow0$ transition is $n_{cr}\sim1500$ cm$^{-3}$), or will
have been dissociated (Keene et al. 1996).  It has been argued
(Stark et al. 1997a) that such ``photon dominated regions'' or
``photodissociation regions'' (PDRs) are more extensive in metal-poor
environments (as the shortage of dust exposes more CO molecules to the
dissociating UV radiation), resulting in a higher ${\rm C/CO}$
ratio. Because the Galactic Center has higher metallicity than the
galactic disk (e.g., Afflerbach, Churchwell, \& Werner 1997), studying
the C/CO ratio there helps in deriving a possible relationship between
metallicity and $\rm{I}_{[\rm{C I}]}/\rm{I}_{^{12}CO}$ (Bolatto et
al. 2000). Such a relationship may be a valuable tool for estimating the
metallicity of distant galaxies.

Past observations of the Galactic Center in [\ion{C}{1}] include those of
White \& Padman (1991), Serabyn et al. (1994), and Jaffe, Plume, \& Evans 
(1996). Serabyn et al. present high resolution observations of a part 
of the Galactic Center region. Jaffe et al.(1996) find that 
[\ion{C}{1}] has very similar distribution (in longitude and velocity) 
to emission in CS and $^{13}$CO. They report that the mean [\ion{C}{1}]
properties of the Galactic Center gas show little change with position.  

\section{Observations}
\label{s:obs} 

The low humidity, high atmospheric stability and thin troposphere at
the South Pole make it an excellent site for submillimeter astronomy
(Chamberlin, Lane, \& Stark 1997, Lane 1998). AST/RO is a general-purpose
telescope built to take advantage of this site. It is a 1.7~m
diameter, offset Gregorian telescope currently capable of observing at
wavelengths between 350~$\mu$m and 1.3~mm (Stark et al. 1997b, Stark
et al. 2000).

Using AST/RO's quasi-optical SIS receiver (Engargiola, Zmuidzinas, 
\& Lo 1994) with
a receiver noise temperature of 180~K, we observed the fine structure
[\ion{C}{1}] $(^{3}P_{1}\rightarrow^{3}P_{0})$ neutral carbon
transition at 492.1607 GHz during the austral winter of 1995. The
telescope beam has HPBW$\sim 2^{'}$ at this frequency. The absolute
pointing error was less than half a beamwidth. The backend was
a 1.1 GHz wide acousto-optical spectrometer (Schieder, Tolls, \&
Winnewisser 1989), with 1.07 MHz wide (0.65 km s$^{-1}$) channels and
a channel spacing of 674 kHz. The
observations were made using position switching 90 arcminutes away in 
Azimuth (or RA, since they are identical at the South Pole). The data 
were calibrated using the procedure described by Stark et al. (2000).  
The typical calibration uncertainty in 492 GHz data obtained by AST/RO 
is $\lesssim10\%$.
 
A total of 74 sparsely sampled spectra were obtained toward the 
Galactic Center region $|l|\leq 2^\circ$, $|b|\leq 0.1^\circ$.
The spacing between observed spectra was $\Delta l=0.1^\circ$ for 
$|l|\leq 1^\circ$ and $\Delta l = 0.2^\circ$ for 
$1^\circ \leq|l|\leq 2^\circ$ at three latitudes: $b = 0^\circ$
and $b = \pm 0.1^\circ$. A typical spectrum is shown in Figure 
\ref{f:GalcenSpec}, and the $l-v$ diagram of the data set is shown in 
Figure \ref{f:Galcenfinal}. Typical rms noise was 0.1~K in 1.07~MHz 
wide channels. 
 
Neutral carbon emission is detected at all positions in all the molecular 
features of the inner galaxy, including the 300 pc nuclear disk.


\section{Results }
\label{s:results}

To compare the molecular and atomic phases of the PDR we have
extracted $^{12}$CO and $^{13}$CO $J=1\rightarrow0$ spectra from the
Bally et al. (1987) survey, which have the same positions and angular
resolution as the [\ion{C}{1}] data. These CO data are shown
in Figures \ref{f:Galcen12CO} and \ref{f:Galcen13CO} (contours) 
together with the [\ion{C}{1}] data (color scale).
 
The rectangular regions shown in Figure \ref{f:Galcenfinal} were chosen 
to be unconfused samples of emission either from Galactic Center 
molecular material or from foreground clouds. The 
$\rm{I}_{[\rm{CI}]}/\rm{I}_{^{12}CO}$ ratio is shown next to each region,
with the corresponding $\rm{I}_{[\rm{CI}]}/\rm{I}_{^{13}CO}$ ratio in 
parentheses. These ratios are summarized in Table \ref{t:CI/COratios}.
 
Although the Galactic Center material has uniformly higher metallicity
than the outer Galaxy, it varies widely in density, temperature, and
optical depth. For example, the diffuse nuclear disk material is a
non-cloudy (i.e. bound to the galaxy as a whole and not to any small 
region), pervasive medium with moderate density ($n\sim1000$ cm$^{-3}$)
and temperature (Stark et al. 1989), whereas Sgr B2 is among the largest,
hottest, and densest clouds in the Galaxy. Nevertheless, the
$\rm{I}_{[\rm{CI}]}/\rm{I}_{^{12}CO}$ ratio is uniformly $\sim0.1$ in
our sample of Galactic Center regions (Figure \ref{f:ratioplots}). 

Kaufman et al. (1999) compute this   
$\rm{I}_{[\rm{CI}]}/\rm{I}_{^{12}CO}$ ratio for PDR models using a 
wide range of density and incident FUV (far ultraviolet) radiation 
field. Using Kaufman's results, our observed ratio (a conversion 
factor of 78 is needed to convert from our integrated 
intensity units of K km s$^{-1}$ to erg s$^{-1}$ cm$^{-2}$ sr$^{-1}$)
corresponds to a density $\sim 3\times10^{4}$ to $ 2\times10^{5}$ 
cm$^{-3}$ and does not constrain the FUV radiation field. This is 
unsurprising since the column density of [\ion{C}{1}] is expected to 
be relatively independent of cloud density and the strength of FUV 
radiation; the radiation field simply determines the depth at which 
the C$^{+}$/CI/CO transition occurs (Hollenbach, Takahashi, \& Tielens 
1991). 

In contrast to the Galactic center material, the foreground regions have
$\rm{I}_{[\rm{CI}]}/\rm{I}_{^{12}CO}$ ratio $\sim0.35$. 
Unlike $\rm{I}_{[\rm{CI}]}/\rm{I}_{^{12}CO}$, 
the $\rm{I}_{[\rm{CI}]}/\rm{I}_{^{13}CO}$ ratio is essentially constant 
over all features of both the inner and outer galaxy. This is in agreement 
with the results of Keene et al. (1996). 
Comparison between the $^{12}$CO and $^{13}$CO spectra (Figure
\ref{f:selfspectra}) shows that the $^{12}$CO
$(J=1\rightarrow0)$ emission is likely 
absorbed within the foreground
clouds, thus explaining their anomalously high 
$\rm{I}_{[\rm{CI}]}/\rm{I}_{^{12}CO}$ ratios. 
This can also be appreciated in Figure \ref{f:ratioplots}.
The spectra corresponding to foreground regions (red symbols) show
anomalously high $\rm{I}_{[\rm{CI}]}/\rm{I}_{^{12}CO}$ and
$\rm{I}_{^{13}CO}/\rm{I}_{^{12}CO}$ ratios, but standard
$\rm{I}_{[\rm{CI}]}/\rm{I}_{^{13}CO}$ ratios. Consequently, [\ion{C}{1}]
appears to be at least as optically thin as $^{13}$CO in all Galactic 
Center and foreground features. The values for the different 
ratios, averaged only over our sample of unconfused Galactic Center
clouds, are $\rm{I}_{[\rm{CI}]}/\rm{I}_{^{12}CO}\approx0.08\pm0.01$,
$\rm{I}_{[\rm{CI}]}/\rm{I}_{^{13}CO}\approx0.75\pm0.20$, and
$\rm{I}_{^{13}CO}/\rm{I}_{^{12}CO}\approx0.11\pm0.01$. These ratios
are broadly in agreement with the ratios reported by Jaffe et al.(1996).

How constant are these ratios throughout the Galactic Center region?
We have performed the following analysis on the entire $^{12}$CO,
$^{13}$CO, and [\ion{C}{1}] datasets: each position observed has been
decomposed into 4 km s$^{-1}$ wide channels between $-$200 and 200 km
s$^{-1}$ LSR velocity, and the integrated intensity has been
calculated for each bin. We have then examined the
$\rm{I}_{[\rm{CI}]}/\rm{I}_{^{12}CO}$,
$\rm{I}_{[\rm{CI}]}/\rm{I}_{^{13}CO}$, and
$\rm{I}_{[\rm^{13}CO]}/\rm{I}_{^{12}CO}$ intensity ratios as a 
function of the corresponding $\rm{I}_{^{12}CO}$, $\rm{I}_{^{13}CO}$, 
and $\rm{I}_{^{12}CO}$ intensities. These ratios are essentially flat
across more than an order of magnitude in integrated
intensity. The average ratios for the complete dataset are
$\rm{I}_{[\rm{CI}]}/\rm{I}_{^{12}CO}\approx0.08\pm0.01$,
$\rm{I}_{[\rm{CI}]}/\rm{I}_{^{13}CO}\approx0.59\pm0.02$, and
$\rm{I}_{^{13}CO}/\rm{I}_{^{12}CO}\approx0.10\pm0.01$, basically
indistinguishable from the average ratios obtained for our sample of
unconfused Galactic Center regions (Figure \ref{f:ratioplots}). While 
the plot for $\rm{I}_{[\rm{CI}]}/\rm{I}_{^{13}CO}$ has some
outlying points, this may not be significant as we have a small sample.







\section{Summary and Conclusions}
\label{s:conclude}
We have presented a sparsely sampled [\ion{C}{1}] map of the Galactic
Center region. The corresponding $l-v$ diagram shows that neutral
carbon emission is detected in all the molecular features of the inner
galaxy.  Despite large variations in the environments and physical 
conditions throughout this region, the integrated intensity ratios
$\rm{I}_{[\rm{CI}]}/\rm{I}_{^{12}CO}$ and
$\rm{I}_{[\rm{CI}]}/\rm{I}_{^{13}CO}$ remain remarkably constant.
For the material identified as foreground, we find that the  
$\rm{I}_{[\rm{CI}]}/\rm{I}_{^{12}CO}$ ratio is clearly higher than in
the material unambiguously identified with the Galactic Center. This 
effect we attribute to the absorption of $^{12}$CO emission within the 
colder foreground material because: 
1) the $\rm{I}_{[\rm{CI}]}/\rm{I}_{^{13}CO}$ ratio is identical
for Galactic Center and foreground clouds, and 2) the individual 
$^{12}$CO spectra show depressions at the relevant velocities.

It is remarkable that the $\rm{I}_{[\rm{CI}]}/\rm{I}_{^{13}CO}$ ratio
is uniform in clouds with such a wide variety of physical conditions 
and metallicity. This observational fact has as yet no satisfactory
theoretical explanation.



\acknowledgments

This research was supported in part by the National Science 
Foundation under a cooperative agreement with the Center for 
Astrophysical Research in Antarctica (CARA), grant number 
NSF OPP 89-20223. CARA is a National Science Foundation 
Science and Technology Center.



\newpage

\figcaption[fig1.eps]{\label{f:GalcenSpec}
Typical [\ion{C}{1}] spectrum in the Galactic Center region, at 
$l=0.0^\circ$, $b=+0.1^\circ$. The narrow features are Galactic 
Center and foreground clouds. The 300 pc nuclear disk emits weakly 
between -110 and 200 km s$^{-1}$.}

\figcaption[fig2.eps]{\label{f:Galcenfinal} Position-velocity
diagram of [\ion{C}{1}] emission. Unconfused regions defined in Table
\protect\ref{t:CI/COratios} are denoted by boxes. Adjacent to each box
are the corresponding [\ion{C}{1}]/$^{12}$CO intensity ratio with the
[\ion{C}{1}]/$^{13}$CO intensity ratio in parentheses (intensities in K 
km s$^{-1}$). Errors on these values are less than
$10\%$ in all cases.  Note that $\rm I_{[CI]}/I_{^{12}CO}\sim0.1$ for
all Galactic Center region clouds, whereas $\rm
I_{[CI]}/I_{^{12}CO}\sim0.35$ for foreground material.}

\figcaption[fig3.eps]{\label{f:Galcen12CO}
Position-velocity diagram for [\ion{C}{1}] emission (color) with 
$^{12}$CO $J=1\rightarrow0$ contours overlaid (contours are 2, 5, 11, 15, 
20 K km s$^{-1}$).}

\figcaption[fig4.eps]{\label{f:Galcen13CO}
Position-velocity diagram for [\ion{C}{1}] emission (color) with 
$^{13}$CO $J=1\rightarrow0$ contours overlaid (contours are 0.3, 0.8, 
1.5, 2.5, 3.5 K km s$^{-1}$).}

\figcaption[fig5.eps]{\label{f:ratioplots} Plots of integrated
intensity ratios for spectra associated with the features identified 
in Table \protect \ref{t:CI/COratios} (one point per spectrum). Points 
for the
foreground material are shown in red, while blue signifies Galactic
Center clouds. The error bars are $1\sigma$ and include 10\%
calibration uncertainty added in quadrature to statistical errors. The
anomalous ratios evident in the foreground material are likely due to
absorption of $^{12}$CO by cooler foreground material. The median value 
for each ratio, computed using only the Galactic Center emission, is 
indicated by the dashed line.}

\figcaption[fig6.eps]{\label{f:selfspectra}
Comparison of $^{12}$CO, $^{13}$CO, and [\ion{C}{1}] spectra 
at a typical position with foreground emission 
($l=-0.8^\circ$, $b=0^\circ$). The $^{12}$CO shows indications of 
absorption at the foreground velocities (-2.5 to 5 km s$^{-1}$).}



\newpage

\begin{deluxetable}{lcccc}
\tablewidth{0pc}
\footnotesize
\tablecaption{[\ion{C}{1}]/CO ratios at the Galactic Center}
\tablehead{
\multicolumn{1}{c}{Region} & $l$ \tablenotemark{a}& V$_{\rm LSR}$ & 
$\rm I_{[CI]}/I_{^{12}\rm{CO}}$ \tablenotemark{b} & 
$\rm I_{[CI]}/I_{^{13}\rm{CO}}$ \tablenotemark{c} \\ 
& (deg) & (km s$^{-1}$) \\
}
\startdata
135 km s$^{-1}$ arm & $-0.6$ $\rm{to}$ $-0.9$ & 
$115$ $\rm{to}$ $160$ & $0.08 \pm 0.004$ & $0.88\pm0.08$ \\ 
 
Sgr A               & $0.1$ $\rm{to}$ $-0.3$ & 
$50$ $\rm{to}$ $125$ & $0.10 \pm 0.008$ & $0.93\pm0.02$ \\
 
Diffuse Nuclear Disk Region 1 & $1.8$ $\rm{to}$ $1.6$ & 
$5$ $\rm{to}$ $50$ & $0.11 \pm 0.012$ & $ 0.92\pm0.05$ \\
 
Diffuse Nuclear Disk Region 2 & $1.2$ $\rm{to}$ $0.9$ & 
$5$ $\rm{to}$ $40$ & $0.11 \pm 0.004$ & $0.95\pm0.07 $\\
 
Diffuse Nuclear Disk Region 3 & $0.9$ $\rm{to}$ $0.5$ & 
$-90$ $\rm{to}$ $-65$ & $0.12 \pm 0.005$ & $1.25\pm0.15$\\
 
3 kpc arm Region 1 & $2.0$ &
$-65$ $\rm{to}$ $-35$ & $0.05 \pm 0.006$ & $1.06\pm0.54$ \\
 
3 kpc arm Region 2 & $-1.2$ $\rm{to}$ $-1.4$ & 
$-75$ $\rm{to}$ $-35$ & $0.20 \pm 0.034$ & $0.78\pm0.20$\\
 
3 kpc arm Region 3 & $-2.0$ & 
$-75$ $\rm{to}$ $-35$ & $0.11 \pm 0.041$ & $0.39\pm0.06$\\

Sgr B & $0.7$ $\rm{to}$ $0.3$ & 
$10$ $\rm{to}$ $55$ & $0.11 \pm 0.001$ & $0.79\pm0.01$ \\
 
300 pc nuclear disk Region 1 & $1.4$ $\rm{to}$ $1.2$ & 
$60$ $\rm{to}$ $110$ & $0.07 \pm 0.001$ & $0.74\pm0.01$\\
 
300 pc nuclear disk Region 2 & $-0.4$ $\rm{to}$ $-0.8$ &
$-160$ $\rm{to}$ $-100$ & $0.14 \pm 0.007$ & $1.05\pm0.03$ \\
 
Foreground 1 & $-0.6$ $\rm{to}$ $-0.9$ & 
$-2.5$ $\rm{to}$ $5$ & $0.43 \pm 0.020$ & $0.92\pm0.02$ \\
 
Foreground 2 & $-2.0$ & 
$5$ $\rm{to}$ $12.5$ & $0.30 \pm 0.039$ & $0.57\pm0.02$\\

\enddata
\label{t:CI/COratios}
\tablenotetext{a}{The Galactic latitude range is $b=-0.1^\circ$ to 
$0.1^\circ$ in all cases.}
\tablenotetext{b}{Ratio of integrated intensities (intensities in K km 
s$^{-1}$, to convert to a cooling ratio multiply by 78).}
\tablenotetext{c}{Ratio of integrated intensities (intensities in K km 
s$^{-1}$, to convert to a cooling ratio multiply by 89).}
\end{deluxetable}

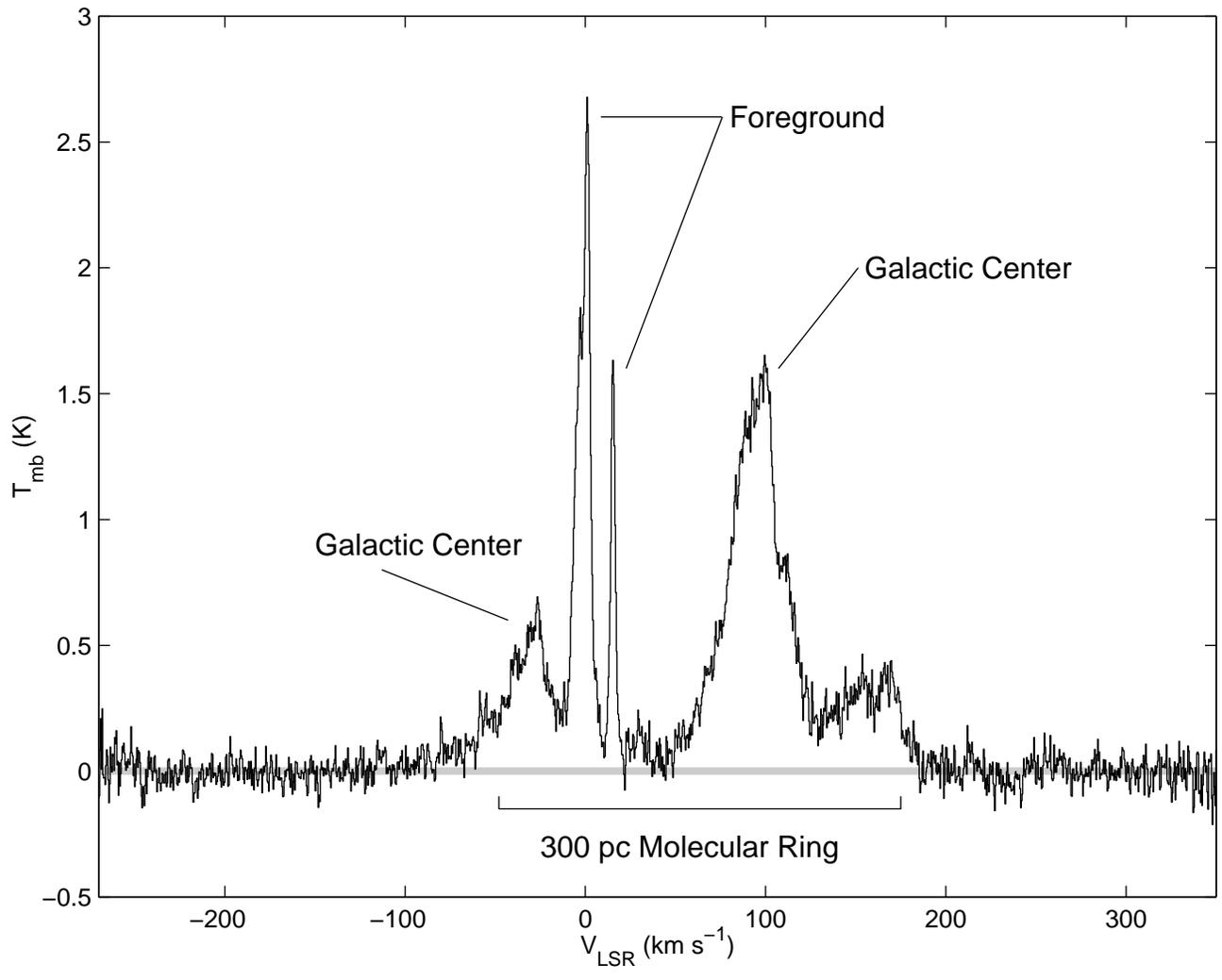
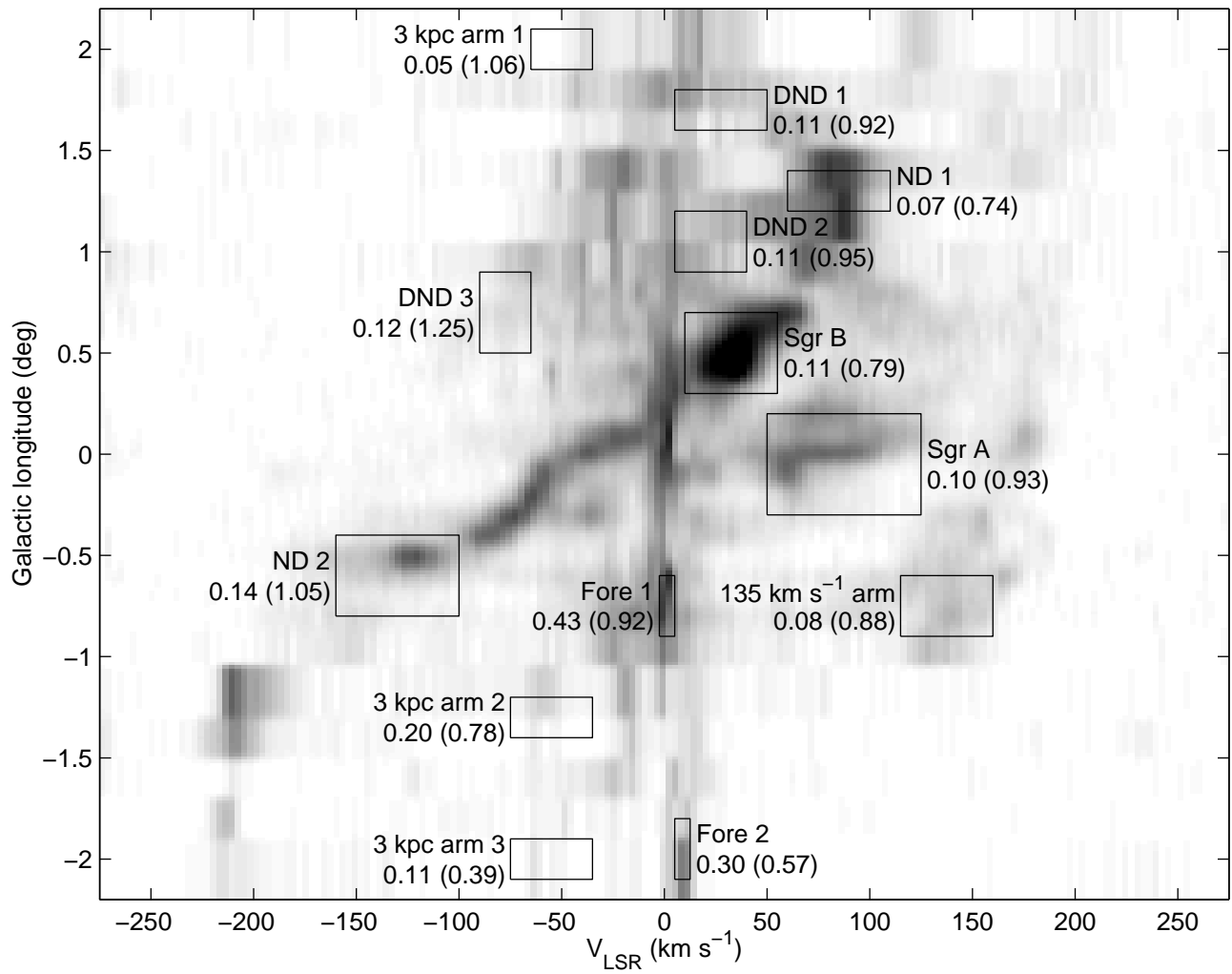
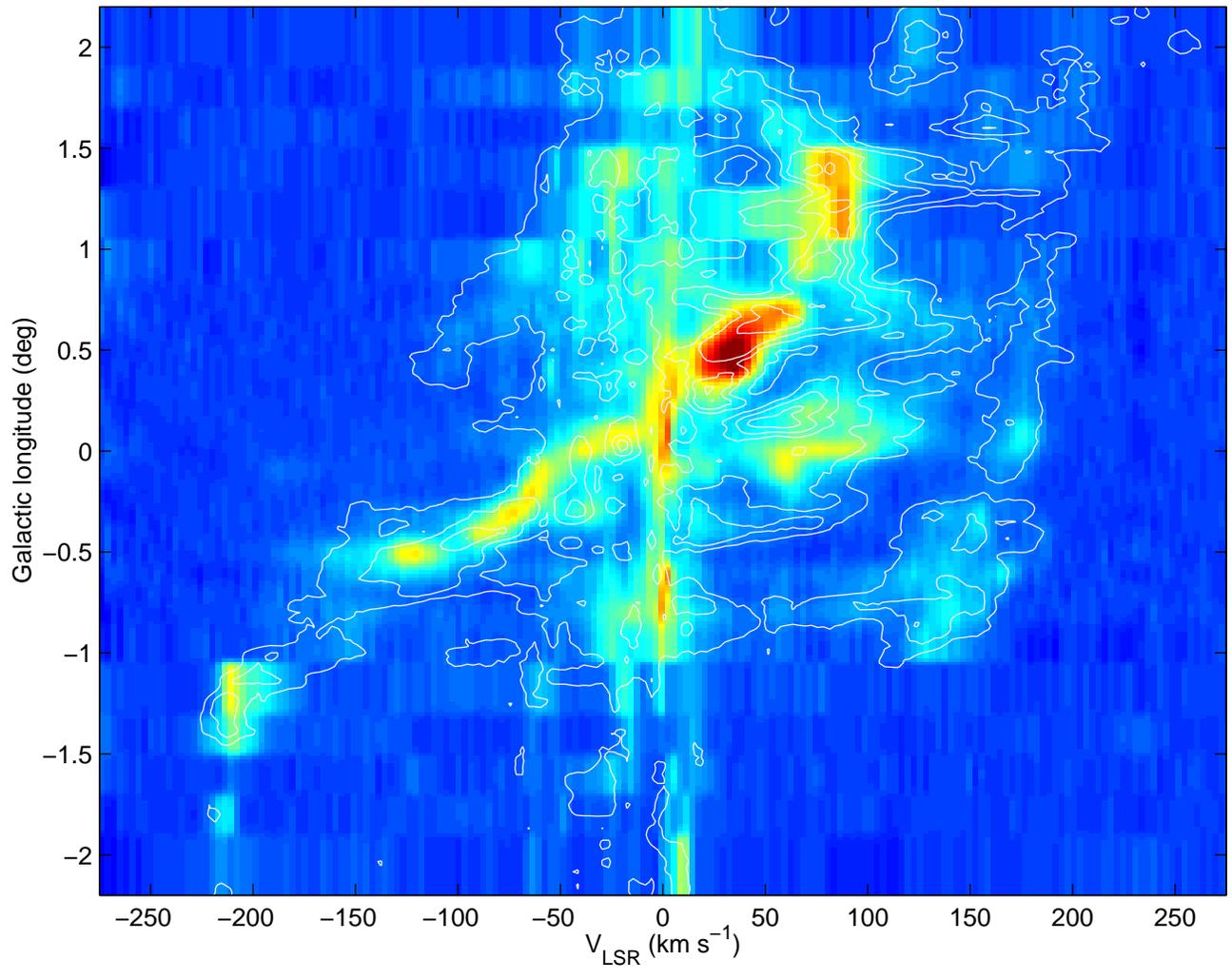
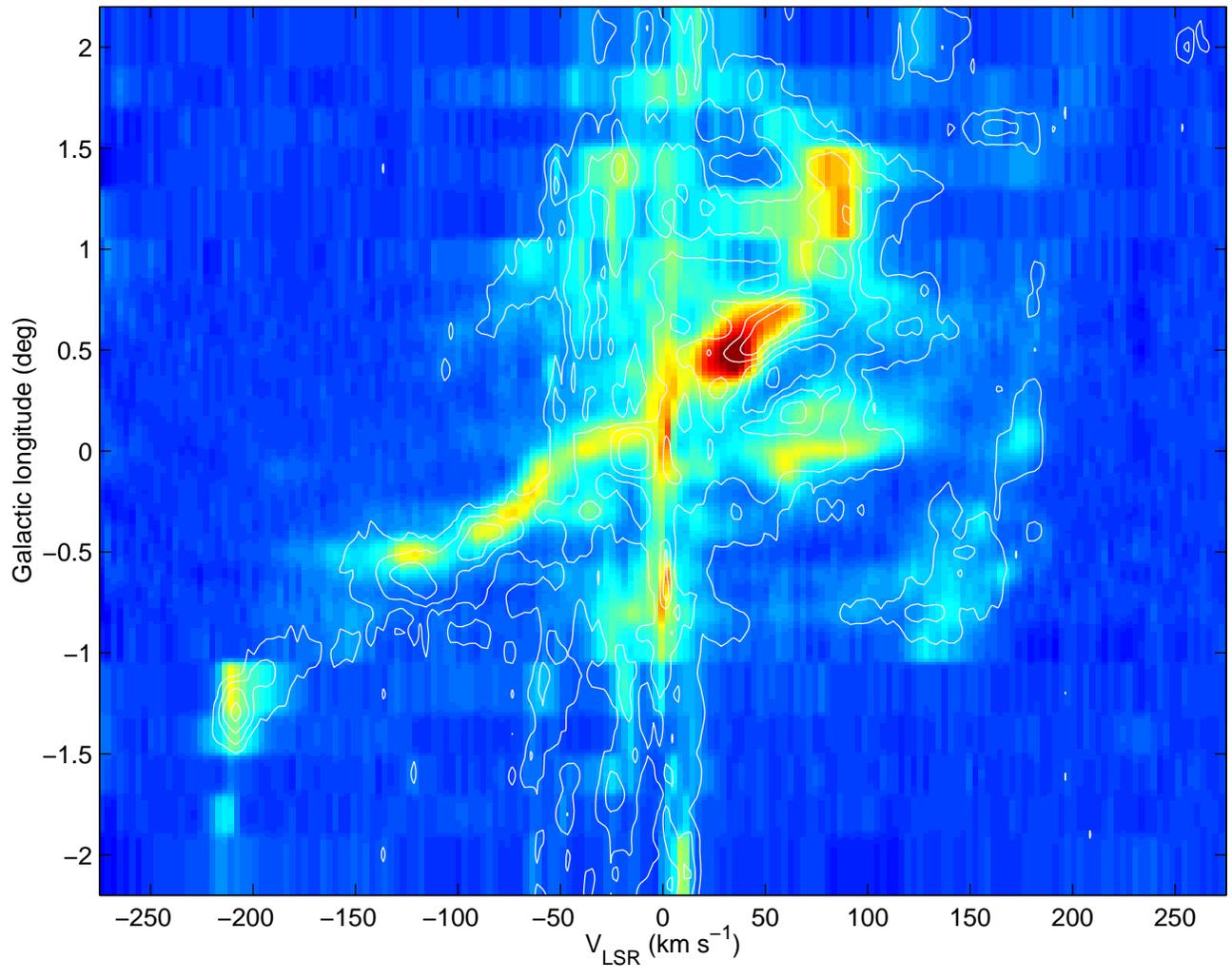
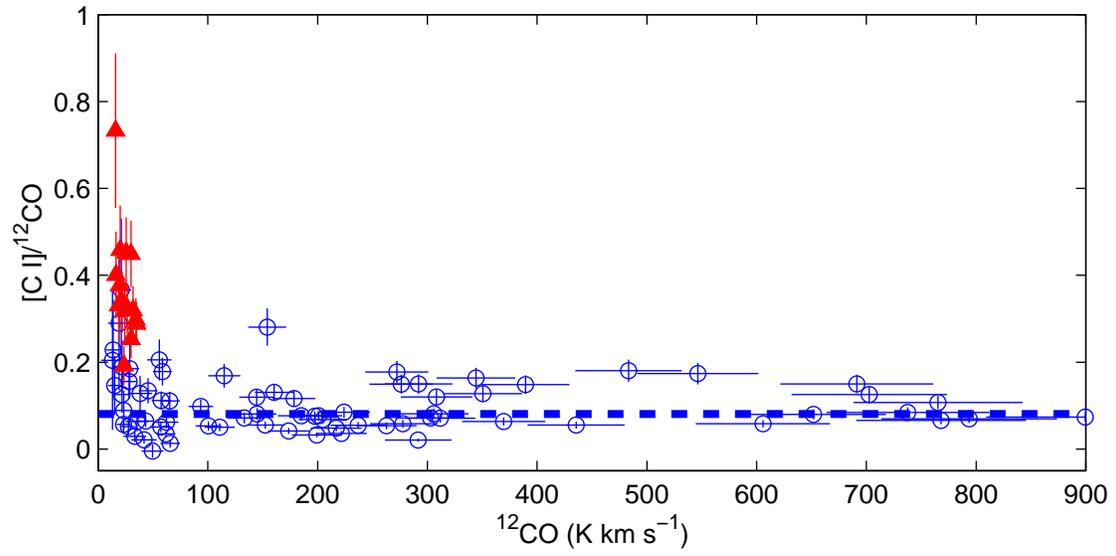
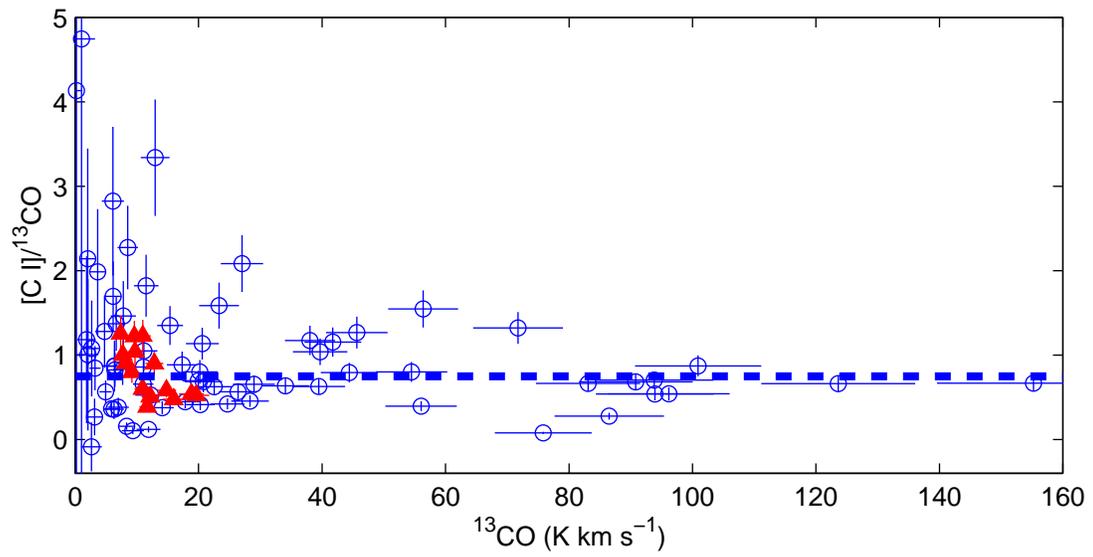
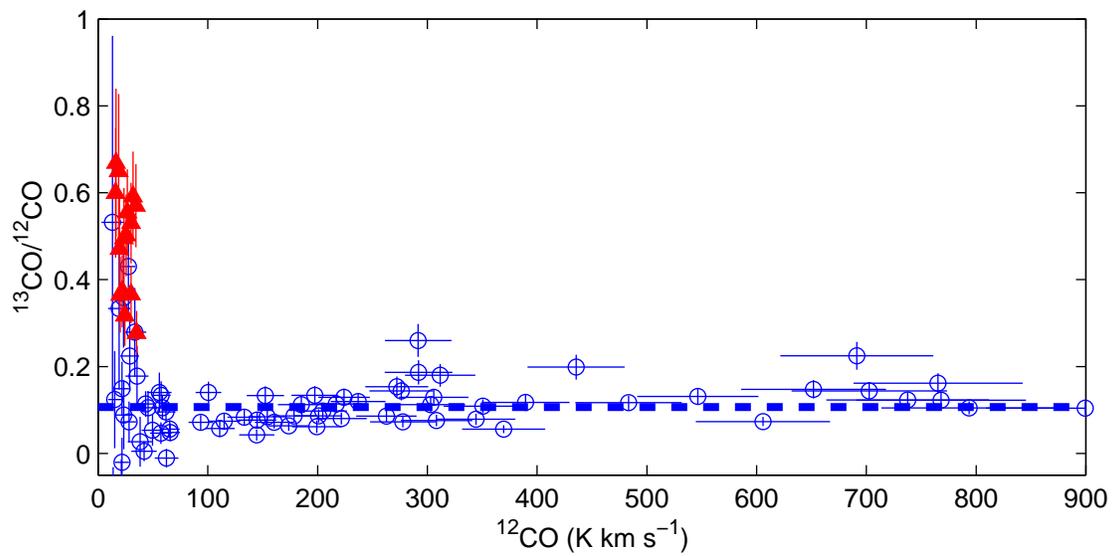
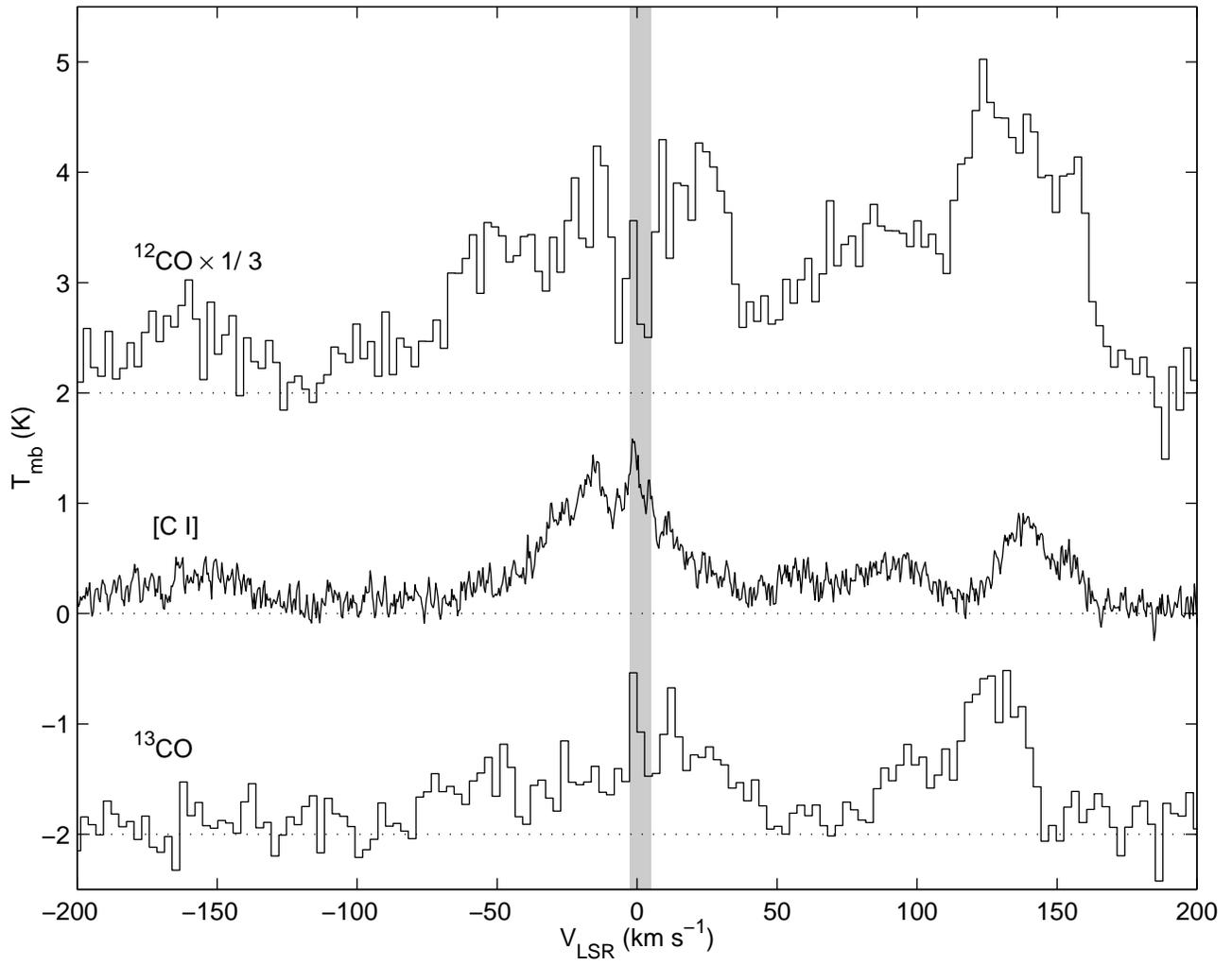
\end{document}